\documentclass[aps,prd,twocolumn,showpacs,preprintnumbers,amsmath,amssymb]{revtex4}   

\usepackage{graphicx}
\usepackage{dcolumn}
\usepackage{bm}
\usepackage{amsfonts}
\usepackage{amsmath,amssymb}

\newcommand {\beq} {\begin{equation}}
\newcommand {\eeq} {\end{equation}}
\newcommand {\beqa}{\begin{eqnarray}}
\newcommand {\eeqa}{\end{eqnarray}}
\newcommand {\del} {\partial}

\newcommand {\ee}{\mbox{e}}

\begin{document}


\title{
New insights into the problem with a singular drift term\\
in the complex Langevin method
}
 
\author{Jun Nishimura$^{1,2}$}
\email{jnishi@post.kek.jp}
\author{Shinji Shimasaki$^{1}$}
\email{simasaki@post.kek.jp}


\affiliation{
$^{1}$KEK Theory Center, High Energy Accelerator Research Organization, 
		Tsukuba 305-0801, Japan \\
$^{2}$Department of Particle and Nuclear Physics,
School of High Energy Accelerator Science,
Graduate University for Advanced Studies (SOKENDAI),
Tsukuba 305-0801, Japan\\
}

\date{April, 2015; preprint: KEK-TH-1816
}


\begin{abstract}
The complex Langevin method aims at performing
path integral with a complex action
numerically based on complexification of the original
real dynamical variables.
One of the poorly understood issues concerns
occasional failure in the presence of
logarithmic singularities in the action,
which appear, for instance, from the fermion determinant
in finite density QCD.
We point out that the failure should be attributed 
to the breakdown of the relation between the complex
weight that satisfies the Fokker-Planck equation
and the probability distribution associated with the stochastic process.
In fact, this problem can occur, in general, when
the stochastic process involves a singular drift term.
We show, however, in a simple example that there exists a parameter region
in which the method works, 
although the standard reweighting method is hardly applicable.
%
%
\end{abstract}


\maketitle


\section{Introduction}

The path integral formulation plays an important role
in nonperturbative studies of quantum field theories
due to the possibility of using Monte Carlo methods.
The basic idea is to generate field configurations
with a probability $\ee^{-S}$ and to 
evaluate the path integral by taking a statistical average.
It is not straightforward, however, to apply such an
approach to cases with a complex action $S$
since one can no longer view $\ee^{-S}$ as the probability distribution.
This ``complex action problem'' occurs, for instance,
in QCD at finite density or with a theta term,
in Chern-Simons gauge theories, and in chiral gauge theories.
It also appears in supersymmetric gauge theories and
matrix models, which are relevant in
nonperturbative studies of superstring theory.

The complex Langevin method (CLM) \cite{Parisi:1984cs,Klauder:1983sp}
attempts to solve this problem
by extending the idea of stochastic quantization 
\cite{Parisi:1980ys}
for ordinary systems with a real action
to the case with a complex action.
This necessarily requires
complexifying the real dynamical variables that
appear in the original path integral.
A stochastic process for the complexified variables
is defined by the Langevin equation with the complex action,
and expectation values in the original path integral
are calculated from an average of
corresponding quantities over the stochastic process.
See Ref.~\cite{Damgaard:1987rr} for a pedagogical review 
on this method.

One of the recent developments in this method is
the clarification of
a necessary condition for convergence to 
correct results \cite{Aarts:2011ax}.
%
In fact, there is a lot of freedom in 
defining the stochastic process that corresponds formally to
the original path integral.
By using this freedom, one can try to satisfy the conditions 
for convergence.
The so-called gauge cooling \cite{Seiler:2012wz}
is a new technique of this
kind, which has made it possible to study
finite density QCD in the high temperature (deconfined) 
phase \cite{Sexty:2013ica}.

Despite these developments, there 
is still a
puzzle concerning the treatment of logarithmic singularities 
in the action \cite{Mollgaard:2013qra,Mollgaard:2014mga,Greensite:2014cxa}.
For instance, the effective action for QCD
involves the logarithm of a determinant,
which represents the effects of dynamical quarks.
At finite density,
the determinant $\Delta$ becomes complex in general, which
causes the complex action problem.
In this case, the effective action has
an ambiguity due to the branch cut of the logarithm;
the phase of the determinant can be defined only modulo $2\pi$.
For instance, one may use the drift term
$\del \log \Delta = \Delta^{-1}\del \Delta$ 
in the complex Langevin equation,
which corresponds to regarding the $-\log \Delta$ term in the action
as a multi-valued function.
It has been found, however, that the CLM with this prescription
can give wrong results in a simplified model when 
the phase of the determinant rotates frequently
during the stochastic 
process \cite{Mollgaard:2013qra}.
The wrong results turn out to be
close to (but not equal to)
the results obtained for the phase-quenched model,
in which the phase of the determinant is quenched.
This problem has not shown up yet
in recent QCD simulations at finite density \cite{Aarts:2014bwa,Sexty:2014dxa}.
%

Here we provide new insights into
this issue from the viewpoint of the Fokker-Planck (FP)
equation.
We start with simple examples and discuss 
more general cases towards the end.


\section{One-variable case}

\subsection{Formulation of the CLM}
\label{sec:model-method}

Let us consider a simple example
defined by the partition function
\begin{eqnarray}
Z = \int dx \, w(x) \ ,\quad
w(x) = (x+i\alpha)^p \, \ee^{-x^2/2} \ ,
\label{part-1var}
\end{eqnarray} 
where $x$ is a real variable and 
$\alpha$ and $p$ are real parameters.
For $\alpha \neq 0$ and $p\neq 0$,
the weight $w(x)$ is complex,
and the idea of important sampling cannot
be applied to (\ref{part-1var})
by regarding $w(x)$ as the Boltzmann weight.
%

Following the usual procedure in CLM,
we define the drift term
\begin{eqnarray}
v(x) = w(x)^{-1}\frac{\del w(x)}{\del x} = 
\frac{p}{x+i\alpha} - x  \ ,
\label{def-drift-term}
\end{eqnarray} 
and complexify the variable as 
$x \mapsto z = x + iy$.
The action 
\begin{eqnarray}
S(z) = - \log w(z) = - p \log (z+i\alpha)+ z^2/2
\label{action-1var}
\end{eqnarray} 
after the complexification
has a logarithmic singularity at $z=-i\alpha$ for $p\neq 0$,
which causes the aforementioned ambiguity due to the branch cut.
However, we emphasize that this is not an issue, in general,
because all we need in formulating the CLM, as we show below,
is the single-valuedness of the drift term $v(z)$ after the complexification
and the single-valuedness of the complex weight $w(x)$ as a function of $x$.
These are satisfied in the present case 
even for a non-integer $p$.
%
%
%
%
%
%
%
%
%

The complex Langevin equation
corresponding to the partition function (\ref{part-1var})
can be written as
\begin{eqnarray}
\frac{dz}{dt} 
=  
v(z) + \eta(t)
=  \frac{p}{z+i\alpha} - z + \eta(t)  \ ,
\label{CLE-1var}
\end{eqnarray} 
where 
$\eta(t)$ represents a real Gaussian noise
satisfying $\langle \eta(t) \eta(t') \rangle = 2 \, \delta(t-t')$.
We define the probability distribution 
$P(x,y;t)$ of $x(t)$ and $y(t)$ at the Langevin time $t$.
Its time evolution follows the FP-like equation
\begin{eqnarray}
\frac{\del}{\del t} P
&=&  L \, P 
\label{FP-like-general}
\\
&=& 
\frac{\del}{\del x}
\left\{ - 
({\rm Re} \, v)_{z=x+iy}
+ \frac{\del}{\del x}\right \}  P
\nonumber
\\
&~& + \frac{\del}{\del y}
\left\{ 
- 
({\rm Im}  \, v)_{z=x+iy} P
\right\} \ .
\label{FP-like-1var}
\end{eqnarray} 

The crucial point in the CLM is that
there exists
a complex weight $\rho(x;t)$, which is related to 
the probability distribution $P(x,y;t)$ through
\begin{eqnarray}
\int {\cal O}(x) \rho(x;t) dx
= \int {\cal O}(x+iy) P(x,y;t) dx dy 
\label{rho-P-rel}
\end{eqnarray} 
under certain conditions \cite{Aarts:2011ax},
where ${\cal O}(x)$ are observables that
admit holomorphic extension to ${\cal O}(x+iy)$.
The evolution of 
$\rho$ follows
the usual FP equation
\begin{eqnarray}
\frac{\del}{\del t} \rho
&=&  L_0 \rho  \ ,  
\label{FPeq-general}
\\
L_0 &=&  \frac{\del}{\del x}
\left( - v(x) + \frac{\del}{\del x}\right )  \ ,
\label{FP-1var}
\end{eqnarray} 
which has a time-independent solution
$\rho(x;t) \propto w(x)$, 
with $w(x)$ given in (\ref{part-1var})
since it is annihilated by
the operator in 
parenthesis in (\ref{FP-1var}).
%
%
Thus, the necessary and sufficient conditions
for being able to calculate the expectation value of ${\cal O}$
with respect to (\ref{part-1var}) by the CLM are:
\begin{enumerate}
\item The relation (\ref{rho-P-rel}) between $\rho$ and $P$ holds.
\item The solution $\rho(x;t)$ 
of
the FP equation (\ref{FPeq-general})
asymptotes to 
$w(x)$ as $t\rightarrow \infty$ up to some constant factor.
\end{enumerate}

As possible observables in the present 
example (\ref{part-1var}), we consider 
${\cal O} = x^{k}$, where $k$ is a positive integer.
Assuming the ergodicity of the stochastic process,
the right-hand side of (\ref{rho-P-rel}) at $t=\infty$
can be evaluated by taking the time average
of $z(t)^k$, where $z(t)$ 
is obtained by solving (\ref{CLE-1var}).
We find numerically that 
this method gives correct results only
for sufficiently large $|\alpha|$ for each $p$.
In what follows, we clarify the reason 
why it fails at small $|\alpha|$.

\subsection{Spectrum of the ``FP Hamiltonian''}
\label{sec:spectrum}

First we have investigated numerically
the eigenvalue spectrum
of the ``FP Hamiltonian'' $(-L_0)$ defined by (\ref{FP-1var}),
assuming that the complex weight $\rho(x)$ falls off rapidly
as $|x|\rightarrow \infty$.

As is clear from what we wrote above,
we have an eigenfunction $\rho(x)=w(x)$ with zero eigenvalue
for arbitrary $p$ and $\alpha$.
When $p$ is a positive odd integer and $\alpha=0$,
we have 
another zero mode
$\rho(x) = |x|^p \, \ee^{-x^2/2}$.
%
For $p>1$, negative eigenvalues appear in the small $|\alpha|$ region.
(Note that, when $\alpha=0$, we have
an eigenfunction
$\rho(x) = x  \, \ee^{-x^2/2}$,
which corresponds to the smallest eigenvalue
$\lambda = -(p-1)$ for any $p$.)
Thus, we find that 
the desired solution $\rho(x;t) \propto w(x)$
is obtained
in the long-time limit of the FP equation (\ref{FPeq-general})
at arbitrary $\alpha$ for $p<1$
and at sufficiently large $|\alpha|$ for $p>1$.

In the parameter region where we have negative modes,
the complex weight diverges as
$\rho(x;t) \propto e^{|\lambda_{\rm min}|t} \rho_{\rm min}(x)$,
where $\lambda_{\rm min}$ is the smallest eigenvalue and 
$\rho_{\rm min}(x)$ is the corresponding 
eigenfunction.
%
%
%
Clearly this behavior 
is incompatible with
the relation (\ref{rho-P-rel}) considering that
$P(x,y;t) \ge 0$ and 
$ \int dx dy \, P(x,y;t) = 1$.
This implies that the relation (\ref{rho-P-rel})
between $\rho$ and $P$ must be broken 
\emph{at least}
in this region.

Applying this kind of argument to a general multi-variable case, 
we find that the condition like ii) 
in section \ref{sec:model-method} is automatically satisfied
if the condition like i) holds, as far as
the probability distribution $P$ asymptotes to a unique 
function. 

\subsection{The relation between $\rho$ and $P$}

Let us then consider what can go wrong with (\ref{rho-P-rel}).
In the derivation of (\ref{rho-P-rel}) given
in Ref.~\cite{Aarts:2011ax}, the authors use
\begin{eqnarray}
&~& 
\int {\cal O}(x+iy) P(x,y;t) dx dy 
 \nonumber \\
&=& 
\int  {\cal O}(x+iy;t) P(x,y;0) dx dy  \ ,
\label{P-condition}
\end{eqnarray} 
where ${\cal O}(x+iy;t)$ is defined by solving 
\begin{eqnarray}
\frac{d }{dt} {\cal O}(x+iy;t)
&=& L^{\top} {\cal O}(x+iy;t) 
\label{O-evolution}
\end{eqnarray} 
with the initial condition ${\cal O}(x+iy;0) = {\cal O}(x+iy)$.
The symbol $L^{\top}$ represents an operator 
satisfying $\langle L^{\top}f,g \rangle=\langle f,L g \rangle$,
where $\langle f, g \rangle  \equiv \int f(x,y)g(x,y)dxdy$,
assuming that $f$ and $g$ are regular functions with sufficiently
fast fall-off as $|x|,|y|\rightarrow \infty$.
%

In order to prove (\ref{P-condition}), 
they consider \cite{Aarts:2011ax}
\begin{eqnarray}
F(\tau) = \int {\cal O}(x+iy;\tau) P(x,y;t-\tau) dx dy 
\label{interpolate-OP}
\end{eqnarray} 
for $0\le \tau \le t$,
which interpolates both sides of Eq.~(\ref{P-condition}).
Taking the derivative with respect to $\tau$, they get
\begin{eqnarray}
\frac{d}{d\tau}F(\tau) \!\!
&=& \!\!
  \int \Big\{ L^{\top} {\cal O}(x+iy;\tau) 
\Big\} P(x,y;t-\tau) dx dy 
\nonumber
\\
&~& \!\! - \!\! \int {\cal O}(x+iy;\tau) L P(x,y;t-\tau) dx dy  \ .
\label{derivative-F}
\end{eqnarray} 
Naively, the two terms cancel through integrating by parts,
which implies that $F(\tau)$ is independent of $\tau$
and hence (\ref{P-condition}).
In order to justify the partial integration, however,
one should be able to neglect the boundary terms.
This requires that
$P(x,y;t)$ decreases sufficiently fast as
$|x|,|y|\rightarrow \infty$ \cite{Aarts:2011ax}.
In addition to this requirement, one also needs
the holomorphy of the drift and of the observables
to prove (\ref{rho-P-rel}).

\subsection{Diverging boundary terms due to singularity}

In the present example, the fast fall-off
of $P(x,y;t)$
as $|x|,|y|\rightarrow \infty$ is satisfied
due to the $-z$ term in (\ref{CLE-1var}).
However, 
we should be 
careful of the singularity at $(x,y)=(0,-\alpha)$.
In order for the boundary terms to be neglected,
it is required that the limits
\begin{eqnarray}
\lim_{x\rightarrow 0}
\Big[ x  \! \int \!
\frac{{\cal O}(z;\tau)}{| z + i\alpha |^2}
  P(x,y;t-\tau) dy \Big] 
\ ,
\label{boundary-terms1} 
\\
\lim_{y\rightarrow -\alpha}
\Big[(y+\alpha) \! \int \! \frac{ {\cal O}(z;\tau)}{| z + i\alpha |^2}
 P(x,y;t-\tau) dx \Big]
\label{boundary-terms2} 
\end{eqnarray} 
should exist for arbitrary $t$ and $\tau$.
Note, in particular, that ${\cal O}(z;\tau)$ obtained
by solving (\ref{O-evolution}) is highly singular
at $z = - i\alpha$
since the operator $L$ appearing 
on the right-hand side of (\ref{O-evolution})
involves the singularity.
For instance, 
let us take the $n$th derivative of the above expressions
with respect to $\tau$ at $\tau=0$.
Using (\ref{O-evolution}),
we obtain terms such as
\begin{eqnarray}
\lim_{x\rightarrow 0}
\Big[ x  \! \int \!
\frac{(L^{\top})^n  {\cal O}(z)}{| z + i\alpha |^2}
  P(x,y;t) dy \Big]   \ ,
\label{boundary-terms1-limit} 
\end{eqnarray} 
which diverges as
$\sim \frac{1}{|x|^{(2n-1)}}$ for $n \ge 1$
if $P(x,y;t)$ is nonzero at $(x,y)=(0,-\alpha)$.

In order to describe the actual situation,
let us define
the radial distribution
\begin{eqnarray}
\varphi(r)\! =\! \frac{1}{2\pi r} \!\!
\int \!\! P(x,y,\infty)\,  \delta(\sqrt{x^2+(y+\alpha)^2}-r)dxdy
\label{def-f}
\end{eqnarray} 
around the singular point $(x,y)=(0,-\alpha)$.
%
%
For small $|\alpha|$,
we observe that $\varphi(r)\sim r$ at small $r$
as long as $p$ is not very large.
In this case, (\ref{boundary-terms1-limit}) 
still diverges for sufficiently large $n$,
and the relation (\ref{rho-P-rel})
between $\rho$ and $P$ can be violated.
Indeed we find that the CLM yields wrong results in 
such a parameter region.
%
%

\subsection{Results for large $p$}
\label{sec:large-p}

In the partition function (\ref{part-1var}),
it is the prefactor $(x+i\alpha)^p$ that causes
the complex action problem.
In view of this, 
one might
think that
the CLM fails
when the phase of $(z(t)+i\alpha)^p$
rotates frequently during the time evolution from (\ref{CLE-1var}). 
%
%
We find that this is not necessarily the case.
%

\begin{figure}[b]
\begin{center}
\includegraphics[height=6cm]{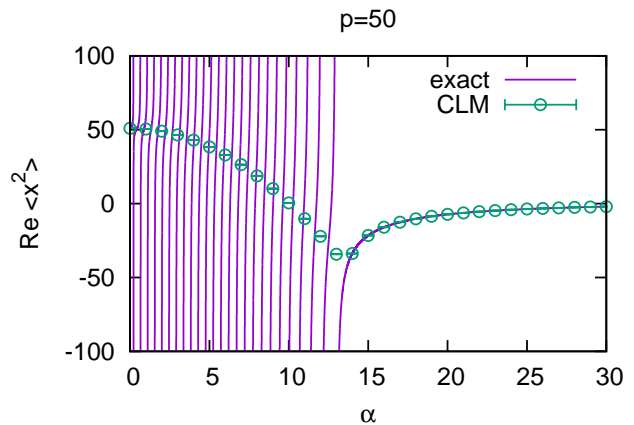}
\end{center}
\caption{
The real part of the expectation value
$\langle x^2 \rangle$
obtained by the CLM (\ref{CLE-1var})
is plotted against $\alpha$ for $p=50$.
The solid line represents the exact result obtained
analytically by the Gaussian integration.
}
\label{ev-var1}
\end{figure}

\begin{figure}[t]
\begin{center}
\includegraphics[height=6cm]{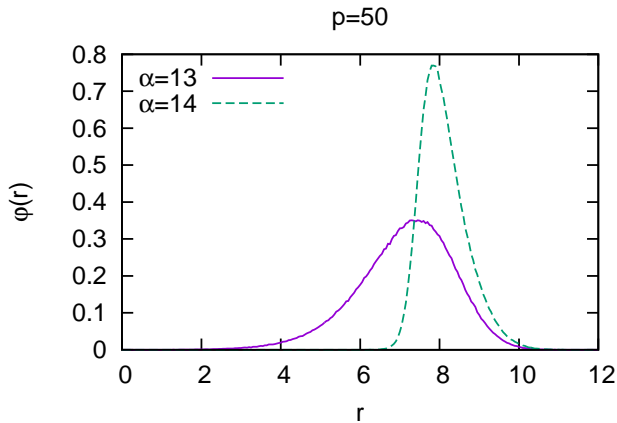}
\end{center}
\caption{
The radial distribution $\varphi(r)$ defined in (\ref{def-f})
is plotted for $p=50$ with
$\alpha=13$ (solid line) and $\alpha=14$ (dashed line).
}
\label{ev-var2}
\end{figure}

In order to demonstrate this point, we present our
results for large $p$.
Figure 1 shows that the CLM
reproduces the exact results for $|\alpha| \gtrsim 14$ at $p=50$.
From Fig.~2, we find for $\alpha=14$ that 
$\varphi(r)=0$ at $r \lesssim 6$,
although we observe that
the phase of $(z(t)+i\alpha)^{50}$ rotates frequently
during the stochastic process.
%
%
%

We would also like to mention that 
the complex action problem is extremely severe 
at $\alpha=14$ and $p=50$.
As a standard quantity that measures
the severeness of the complex action problem,
let us consider
\begin{eqnarray}
R = \left\langle \frac{w(x)}{|w(x)|} \right\rangle_0
=\left\langle \frac{(x+i\alpha)^p}{|x+i\alpha|^p} \right\rangle_0
=\frac{Z}{Z_0}  \ ,
\label{ratio-phase-quenched}
\end{eqnarray} 
where $Z_0$ is the partition function of the phase-quenched model
$Z_0 = \int dx \, |w(x)| $
and 
the expectation value $\langle \ \cdot \  \rangle_0$
is taken 
with respect to it.
In the present case, 
both $Z$ and $Z_0$ can be calculated analytically 
by performing the Gaussian integration.
We find that $R\sim -7.4 \times 10^{-5}$ at $\alpha=14$ and $p=50$.
One can imagine how hard it is to obtain correct results
if one performs a Monte Carlo simulation of the 
phase-quenched model
and applies the standard reweighting formula
to obtain the expectation values
with respect to the original partition function (\ref{part-1var}).
Thus, the advantage of the CLM over the reweighting method
can be appreciated even in this simple one-variable case.


\subsection{Non-logarithmic singularities}
\label{sec:non-log}

Another interesting 
implication of our argument is
that the possible failure of the CLM is not
specific
to logarithmic singularities in the action.
Indeed we have found that the CLM can fail
for the weight $w(x)=e^{-S(x)}$ with the action
\begin{eqnarray}
S(x) = \beta (x+i\alpha)^{-2} + x^2/2 \ .
\label{action-1var-no-log}
\end{eqnarray}
Note that the action $S(z)$ after complexification
does \emph{not} involve a logarithmic singularity,
which means, in particular, that there is no issue 
of ambiguity associated with the branch cut.
Nevertheless, we find that
the CLM fails 
at $|\alpha|\lesssim 1.2$ for $\beta=1$
and at $|\alpha|\lesssim 1.7$ for $\beta=-1$.
%
On the other hand, 
from the studies of the eigenvalue spectrum of $(-L_0)$,
we find that $\rho(x;t) \propto e^{-S(x)}$ is obtained
in the long-time limit of the FP equation 
(\ref{FPeq-general})
for arbitrary $\alpha$ with $\beta=\pm 1$.
Therefore, the failure of the CLM at small $|\alpha|$ should be
attributed to the violation of (\ref{rho-P-rel})
due to the singularity in 
the drift term $v(z)= 2 \beta (z+i\alpha)^{-3} - z$.
This is also confirmed 
from 
the behavior of the radial distribution (\ref{def-f}).

\section{Two-variable case}
\label{sec:two-var}

Our argument applies also to the case with multiple variables.
To make this clear, let us consider a case with two variables
given by
\begin{eqnarray}
Z &=& \int dx_1 dx_2 \, w(x_1,x_2) \ , 
\label{part-2var}
\\
w(x_1,x_2) &=& (x_1 + i  x_2)^p
 \, \ee^{-(x_1)^2/2 - (x_2-\alpha)^2/2}  \ ,
\label{w-2var}
\end{eqnarray} 
where $x_1$ and $x_2$ are real variables.
The parameter $\alpha$ is real, while $p$ is a positive integer.

We have studied numerically
the eigenvalue spectrum of the operator $(-L_0)$,
assuming that the complex weight $\rho(x_1,x_2)$ falls off rapidly
as $(x_1)^2 + (x_2)^2 \rightarrow \infty$.
%
First we obtain the desired zero mode
$\rho(x_1,x_2) = w(x_1,x_2)$
for arbitrary $p$ and $\alpha$.
When $\alpha=0$,
we have another zero mode
$ \rho = |x_1 + i x_2|^p
 \, \ee^{-(x_1)^2/2 - (x_2)^2/2} $ for any $p$.
For $p>1$, negative modes appear at small $|\alpha|$.
(Note that, when $\alpha=0$, we have an eigenfunction
$\rho =
(x_1+i x_2) 
 \, \ee^{-(x_1)^2/2-(x_2)^2/2} $,
which corresponds to the smallest eigenvalue $\lambda = -(p-1)$ for any $p$.)
Thus, we can make an argument analogous to
the one-variable case (\ref{part-1var}).
Indeed we find for $p=1,2,3$ that the CLM with
complexified variables $z_1$ and $z_2$
gives wrong results 
at small $|\alpha|$.
This can be understood from the behavior of
the radial distribution for $r= | z_1 + i z_2|$.
%
%


\section{Implications to finite density QCD}

Let us discuss the implication of our argument
to finite density QCD,
which involves the complex fermion determinant
${\rm det}(D+m)$ in the partition
function,
where $D$ represents the Dirac operator and $m$ is the quark mass.
The determinant can be written as the product
of the eigenvalues $\lambda_k$ of $(D+m)$.
The drift term of the complex Langevin equation
involves $\sum_k  (\lambda_k)^{-1} \del \lambda_k$,
where $\del$ represents the derivative with respect to the 
complexified gauge field.

According to our argument, 
the problem we discussed does not appear
as long as the distribution of $\lambda_k$
is practically zero at the origin,
even if the phase of the fermion determinant rotates frequently
during the stochastic process.
(See our results for large $p$ in the one-variable case.)
This is consistent with 
the results of recent QCD simulations at finite density,
where the distribution of $\lambda_k$ has the desired property
due either to large quark mass \cite{Aarts:2014bwa}
or to high temperature \cite{Sexty:2014dxa}.

On the other hand,
the eigenvalues of $D$ obtained in the CLM
are speculated to accumulate at the origin 
in the chiral limit (corresponding to the $m \rightarrow 0$ limit)
when the chiral symmetry is spontaneously broken \cite{Splittorff:2014zca}.
If true, the CLM will have problems
in that parameter regime unless some new idea is invoked.



\section{Summary}
We have discussed the issue in the CLM 
concerning the logarithmic singularities in the action.
The standard drift term
corresponds to regarding the logarithm
in the action 
as a multi-valued function of 
the complexified variables.
The CLM with this drift term
is known to give wrong results in some cases.
Theoretical understanding of this problem is
important, for instance, 
in applying the method to finite density QCD 
in the low temperature (confined) phase 
with light quark mass.

First we emphasized that
the multi-valuedness of 
logarithmic terms in the action
cannot be considered the cause of the problem 
since one can formulate the 
method without referring to the action,
as we have done in section \ref{sec:model-method}.
This is also indicated 
by the example in section \ref{sec:non-log}.

Rather, the problem should be attributed to
the possible breakdown of the key relation
between $\rho$ and $P$ 
due to the singularities in the
drift term of the complex Langevin equation.
%
In particular, we pointed out that
the relation can be violated due to
the boundary terms appearing from integrating by parts
in proving (\ref{P-condition}),
which diverges
unless $P$ is practically zero around the singularities.
This assertion was supported by
simple examples.
A more quantitative analysis will be reported
in the forthcoming publication. 


The ``FP Hamiltonian'' can have negative modes
only if the key relation between $\rho$ and $P$
is violated.
Note, however, that the key relation can be 
violated even if the ``FP Hamiltonian'' does not
have negative modes.
Hence,
the appearance
of the negative modes should be regarded merely
as an indicator of the violation of the key relation,
the latter being the cause of the problem.


To conclude, we hope
that the new insights gained in this work
will be useful in developing the method
further
in cases with singularities in the drift term.

%
%


  \begin{center}
  {\bf\small ACKNOWLEDGMENTS}
  \end{center}

The authors would like to 
thank D.~Sexty for valuable discussions.
The work of J.~N.\ was supported in part by Grant-in-Aid 
for Scientific Research (No.\ 23244057)
from Japan Society for the Promotion of Science.




\begin{thebibliography}{99}


\bibitem{Parisi:1984cs} 
  G.~Parisi,
  Phys.\ Lett.\ B {\bf 131}, 393 (1983).
\bibitem{Klauder:1983sp}
  J.~R.~Klauder,
  Phys.\ Rev.\ A {\bf 29}, 2036 (1984).

\bibitem{Parisi:1980ys} 
  G.~Parisi and Y.~s.~Wu,
  Sci.\ Sin.\  {\bf 24}, 483 (1981).


\bibitem{Damgaard:1987rr} 
  P.~H.~Damgaard and H.~Huffel,
  Phys.\ Rept.\  {\bf 152}, 227 (1987).

\bibitem{Aarts:2011ax} 
  G.~Aarts, F.~A.~James, E.~Seiler and I.~O.~Stamatescu,
  Eur.\ Phys.\ J.\ C {\bf 71}, 1756 (2011)
  [arXiv:1101.3270 [hep-lat]].

\bibitem{Seiler:2012wz} 
  E.~Seiler, D.~Sexty and I.~O.~Stamatescu,
  Phys.\ Lett.\ B {\bf 723}, 213 (2013)
  [arXiv:1211.3709 [hep-lat]].

\bibitem{Sexty:2013ica} 
  D.~Sexty,
  Phys.\ Lett.\ B {\bf 729}, 108 (2014)
  [arXiv:1307.7748 [hep-lat]].


\bibitem{Mollgaard:2013qra} 
  A.~Mollgaard and K.~Splittorff,
  Phys.\ Rev.\ D {\bf 88}, no. 11, 116007 (2013)
  [arXiv:1309.4335 [hep-lat]].

\bibitem{Mollgaard:2014mga} 
  A.~Mollgaard and K.~Splittorff,
  Phys.\ Rev.\ D {\bf 91}, no. 3, 036007 (2015)
  [arXiv:1412.2729 [hep-lat]].

\bibitem{Greensite:2014cxa} 
  J.~Greensite,
  Phys.\ Rev.\ D {\bf 90}, no. 11, 114507 (2014)
  [arXiv:1406.4558 [hep-lat]].



\bibitem{Aarts:2014bwa} 
  G.~Aarts, E.~Seiler, D.~Sexty and I.~O.~Stamatescu,
  Phys.\ Rev.\ D {\bf 90}, no. 11, 114505 (2014)
  [arXiv:1408.3770 [hep-lat]].

\bibitem{Sexty:2014dxa} 
  D.~Sexty,
  arXiv:1410.8813 [hep-lat].






%




\bibitem{Splittorff:2014zca} 
  K.~Splittorff,
  Phys.\ Rev.\ D {\bf 91}, no. 3, 034507 (2015)
  [arXiv:1412.0502 [hep-lat]].




\end{thebibliography}
\end{document}